\documentclass[letterpaper, conference,10pt] {IEEEtran}
\usepackage{amsmath}

\usepackage{amssymb,amsthm}
\usepackage{bm}
\usepackage{color}
\usepackage{graphicx}
\usepackage{empheq}
\usepackage[linesnumbered,ruled,lined]{algorithm2e}
\usepackage[outdir=./]{epstopdf}
\usepackage{caption}
\usepackage{lipsum}
\usepackage{cite}
\usepackage{multirow}
\usepackage{subcaption}
\usepackage{fancyhdr}
\SetKwInput{Input}{input}
\SetKwInput{Output}{output}

\usepackage{array}
\newcolumntype{L}[1]{>{\raggedright\let\newline\\\arraybackslash\hspace{0pt}}m{#1}}
\newcolumntype{C}[1]{>{\centering\let\newline\\\arraybackslash\hspace{0pt}}m{#1}}
\newcolumntype{R}[1]{>{\raggedleft\let\newline\\\arraybackslash\hspace{0pt}}m{#1}}

\theoremstyle{remark}

\IEEEoverridecommandlockouts
\def\specialpapernotice#1{\if@confmode%
	\def\@specialpapernotice{{\sublargesize\textit{#1}\vspace*{1em}}}%
	\else%
	\def\@specialpapernotice{{\\*[1.5ex]\sublargesize\textit{#1}}\vspace*{-2ex}}%
	\fi}

\begin{document}

	\title{Dirty Paper Coded Rate-Splitting for Non-Orthogonal Unicast and Multicast Transmission with Partial CSIT
   }

		\author{
			
		\IEEEauthorblockN{Yijie Mao and  Bruno Clerckx }
		\IEEEauthorblockA{
		Department of Electrical and Electronic Engineering,	Imperial College London, United Kingdom\\
			Email: \{y.mao16, b.clerckx\}@imperial.ac.uk}

			\thanks{
				This work has been partially supported by the U.K. Engineering and Physical Sciences Research Council (EPSRC) under grant EP/N015312/1, EP/R511547/1.}
		\\[-1 ex]
		{\sublargesize\textit{(Invited Paper)}}	
			\\[-3 ex]		
			}

\maketitle

\thispagestyle{empty}
\pagestyle{empty}
\begin{abstract}
A Non-Orthogonal Unicast and Multicast (NOUM) transmission system allows a multicast stream intended to all receivers to be jointly transmitted with unicast streams in the same time-frequency resource blocks. While the capacity of the two-user multi-antenna NOUM with perfect Channel State Information at the Transmitter (CSIT) is known and achieved by Dirty Paper Coding (DPC)-assisted NOUM with Superposition Coding (SC), the capacity and the capacity-achieving strategy of the multi-antenna NOUM with partial CSIT remain unknown. In this work, we focus on the partial CSIT setting  and make two major contributions. First, we show that linearly precoded Rate-Splitting (RS), relying on splitting  unicast messages into common and private parts, encoding the common parts together with the multicast message and linearly precoding at the transmitter, can achieve larger rate regions than DPC-assisted NOUM with partial CSIT. Second, we study Dirty Paper Coded Rate-Splitting (DPCRS), that marries RS and DPC. We show that the rate region of DPCRS-assisted NOUM  is enlarged beyond that of conventional DPC-assisted NOUM and that of linearly precoded RS-assisted NOUM with partial CSIT. 
\end{abstract}

\section{Introduction}
\par  Non-Orthogonal Unicast and Multicast (NOUM) transmission, also known as layered division multiplexing (LDM) \cite{LDM2016},  has been gaining increasing attentions recently. It is considered to be a promising solution to support a mixture of unicast and multicast services for future wireless communication networks. Different from conventional approaches where unicast and multicast services are carried out in orthogonal resource blocks,  NOUM  allows each user to receive  a dedicated unicast message  and a multicast message  simultaneously based on Superposition Coding (SC) at the transmitter and Successive Interference Cancellation (SIC) at the receivers.

\par There are two conventional approaches studied in the literature of  NOUM for multi-antenna Broadcast Channels (BC). The first approach is the practical Multi-User Linear Precoding (MU--LP)-assisted NOUM \cite{tervo2017energy,chen2017joint,NOUM2019Gunduz} where the encoded multicast and unicast streams are linearly precoded and superimposed  at the transmitter, each user decodes and removes the multicast stream with the assistance of one layer  SIC before decoding its intended unicast stream. The other approach is the non-linear Dirty Paper Coding (DPC)-assisted NOUM relying on DPC to encode unicast messages and SC to encode multicast messages. It is also  known as ``multi-antenna BC with a common message" in \cite{jindal2004optimal,Weing2006Capacity,Capacity2014Geng}. 
 DPC-assisted NOUM has been shown in \cite{Capacity2014Geng} to  achieve the capacity region of two-user multi-antenna NOUM with perfect Channel State Information at the Transmitter (CSIT). However, when the transmitter only has access to partial Channel State Information (CSI), the capacity and the capacity-achieving strategy of multi-antenna NOUM remain  an open problem. 
 
 
 \par Even in the conventional unicast-only multi-antenna BC with partial CSIT, the capacity  and capacity-achieving scheme are still unknown. Interestingly, we have shown in our most recent work \cite{mao2019beyondDPC} that Dirty Paper Coded Rate-Splitting (DPCRS), that relies on Rate-Splitting (RS) to split  user messages into common and private parts, and DPC to encode the private  parts, enlarges the rate region of conventional DPC  in Multiple-Input Single-Output (MISO) BC with partial CSIT.
Moreover, linearly precoded RS, which has been widely studied in multi-antenna networks \cite{RSintro16bruno,RS2016hamdi,mao2017rate, mao2019maxmin,bruno2019wcl}, is able to achieve larger rate region than DPC in multi-antenna BC with partial CSIT.
The application of linearly precoded RS in multi-antenna NOUM has also been recently studied in \cite{mao2019TCOM}. By splitting the unicast  messages of users into common and private parts, jointly encoding the multicast message and the common parts of the private messages into a super-common stream, linearly precoding super-common stream and private streams, RS-assisted NOUM has been shown to achieve higher spectral and energy efficiencies than the conventional MU--LP-assisted or Non-Orthogonal Multiple Access (NOMA)-assisted strategies  thanks to its robustness and flexibility to manage  interference \cite{mao2019TCOM}.

In this work, we first study the performance of DPC and linearly precoded RS in multi-antenna NOUM with partial CSIT.  We show that linearly precoded RS is able to achieve larger rate regions than DPC-assisted NOUM. 
Motivated by the performance benefits of the linearly precoded RS-assisted NOUM as well as the non-linear DPCRS frameworks in the unicast-only transmission, we further propose a novel DPCRS-assisted NOUM transmission strategy relying on splitting  unicast messages into common and private parts, encoding the private parts by DPC and encoding the common parts together with the multicast message at the transmitter. We show that such DPCRS-assisted NOUM achieves larger rate regions than conventional DPC-assisted NOUM  with partial CSIT.

\section{System Model}
\label{sec: system model}
\par Consider a single-cell downlink transmission, which consists of one  multi-antenna Base Station (BS) equipped with $N_t$ antennas simultaneously serving $K$ single-antenna users, indexed by $\mathcal{K}=\{1,\ldots,K\}$. Hybrid unicast and multicast services are provided in the system.  In each scheduled time frame,  the BS delivers one multicast message $W_0$ to all users and $K$  dedicated unicast messages $W_k, k\in\mathcal{K}$ to the corresponding users. The $K+1$  messages are encoded into the stream vector $\mathbf{s}$ and linearly precoded by the precoding matrix $\mathbf{P}$. The resulting transmit signal is $\mathbf{x}=\mathbf{P}\mathbf{s}$, which is subject to the transmit power constraint $\mathbb{E}\{\Vert\mathbf{x}\Vert^2\}\leq P_t$. Under the assumption that $\mathbb{E}\{\mathbf{s}\mathbf{s}^H\}= \mathbf{I}$, we obtain that $\mathrm{tr}(\mathbf{P}\mathbf{P}^H)=P_t$.
The signal received by user-$k$ is 
\begin{equation}
\label{eq: receivedSignalModel}
	y_k=\mathbf{h}_k^H\mathbf{x}+n_k, \forall k\in \mathcal{K}, 
\end{equation}
where $\mathbf{h}_k\in\mathbb{C}^{{N_t}}$ is the channel between the BS and user-$k$. 
$n_k \sim \mathcal{CN}(0,1)$ is the Additive White Gaussian Noise (AWGN). 
Hence, the transmit Signal-to-Noise Ratio (SNR)  is equal to  $ P_t$.

\subsection{Partial Channel State Information}
\label{sec: channelModel}
\par 
In this work, we assume  the CSI of each user is perfectly known at users (i.e., perfect CSIR) and partially known at the BS (i.e., partial CSIT). The actual CSI known at all users is denoted by $\mathbf{H}=[\mathbf{h}_1,\dots,\mathbf{h}_K]$ and the partial instantaneous channel estimate at the BS is denoted by $\widehat{\mathbf{H}}=[\widehat{\mathbf{h}}_1,\dots,\widehat{\mathbf{h}}_K]$. For a given estimate, the CSIT estimation error is denoted by $\widetilde{\mathbf{H}}=[\widetilde{\mathbf{h}}_1,\dots,\widetilde{\mathbf{h}}_K]$. We have the following relationship:
\begin{equation}
	\mathbf{H}=\widehat{\mathbf{H}}+\widetilde{\mathbf{H}},
\end{equation}
The joint distribution of $\{ \mathbf{H},\widehat{\mathbf{H}}\}$  is assumed to be stationary and ergodic \cite{RS2016hamdi}.   Though ${\mathbf{H}}$ over the entire transmission is unknown at the BS, the conditional density $f_{{\mathbf{H}}|\widehat{\mathbf{H}}}({\mathbf{H}}|\widehat{\mathbf{H}})$ is assumed to be known at the BS. 
 Each element of the $k$th-column of $\widetilde{\mathbf{H}}$ for user-$k$ is characterized  by an independent and identically distributed (i.i.d.) zero-mean complex Gaussian distribution variable with $\mathbb{E}\{\widetilde{\mathbf{h}}_k \widetilde{\mathbf{h}}_k^H\}=\sigma_{e,k}^2\mathbf{I}$. The variance of the error $\sigma_{e,k}^2$ is considered to scale exponentially with SNR as $\sigma_{e,k}^2\sim O(P_t^{-\alpha})$. $\alpha\in[0,\infty)$ is the CSIT quality scaling factor  \cite{AG2015,NJindalMIMO2006,doppler2010Caire,DoF2013SYang,RS2016hamdi}. $\alpha=0$ and  $\alpha=\infty$  stands for partial CSIT with finite precision and  perfect CSIT, respectively.

\subsection{Conventional Dirty Paper Coding-Assisted NOUM}
\label{sec: DPC}

\par Conventional DPC-assisted NOUM  relying on DPC to encode unicast messages and SC to encode multicast messages has been studied in \cite{Weing2006Capacity,Capacity2014Geng} for two-user NOUM with perfect CSIT. In this work, we study its performance in the partial CSIT setting. 

\par At the transmitter, the unicast messages $W_k, k\in\mathcal{K}$ are encoded using DPC based on certain  encoding order $\pi$, where ${\pi}\triangleq[\pi(1),\ldots,\pi(K)]$ is a permutation of \{$1,\ldots,K$\} such that the message $W_{\pi(i)}$ is encoded before $W_{\pi(j)}$ if $i<j$.  The BS  encodes the unicast messages $W_{\pi(1)},\ldots,W_{\pi(K)}$  into a set of symbol streams $s_{\pi(1)},\ldots,s_{\pi(K)}$  and precodes the streams by $\mathbf{p}_{\pi(1)},\ldots,\mathbf{p}_{\pi(K)}$ based on DPC, where $\mathbf{p}_{\pi(k)}\in\mathbb{C}^{N_t}$ is the precoder for $s_{\pi(k)}$. The multicast message $W_0$ is encoded into the multicast stream $s_0$, precoded by $\mathbf{p}_{0}\in\mathbb{C}^{N_t}$ and superimposed on top of the unicast streams.  The resulting  transmit signal is 
\begin{equation}
\mathbf{x}=\mathbf{P}_1\mathbf{s}_1=\mathbf{p}_{0}{s}_{0}+{{\sum_{k\in\mathcal{K}}\mathbf{p}_{\pi(k)}{s}_{\pi(k)}}},
\end{equation}
where $\mathbf{s}_1\triangleq[s_0,s_{\pi(1)},\ldots,s_{\pi(K)}]^T$ and $\mathbf{P}_1\triangleq[\mathbf{p}_0,\mathbf{p}_{\pi(1)},\ldots,\mathbf{p}_{\pi(K)}]$. 

At user side, each user first decodes the multicast stream $s_0$ into $\widehat{W}_0$ by treating the interference from all unicast streams as noise. The instantaneous rate at user-$\pi(k)$ to decode the multicast stream $s_0$ is given as
\begin{equation}
\label{eq: DPC multicast rate}
\resizebox{.18 \textwidth}{!} {$R_{0,\pi(k)}^{\textrm{DPC}}( \mathbf{H},\widehat{\mathbf{H}})=\log_2$}\resizebox{.26\textwidth}{!} {$\left(1+\frac{|{\mathbf{h}}_{\pi(k)}^H\mathbf{p}_{0}|^2}{\sum_{j\in\mathcal{K}}|\mathbf{h}_{\pi(k)}^H\mathbf{p}_{\pi(j)}|^2+1}\right)$}.
\end{equation}
Once the common message $\widehat{W}_0$ is decoded, it is then removed from the received signal by SIC. Assuming perfect SIC, the received signal at user-$\pi(k)$  after removing $\widehat{W}_0$ is
\begin{equation}
\small
y_{\pi(k)}=\widetilde{\mathbf{h}}_{\pi(k)}^H\sum_{i<k}\mathbf{p}_{\pi(i)}s_{\pi(i)}+{\mathbf{h}}_{\pi(k)}^H\sum_{j\geq k}\mathbf{p}_{\pi(j)}s_{\pi(j)}+n_{\pi(k)}.
\end{equation}
Note that since DPC at the BS is implemented based on the  channel estimate $\widehat{\mathbf{H}}$, only  the interference part 
 $\widehat{\mathbf{h}}_{\pi(k)}^H\sum_{i<k}\mathbf{p}_{\pi(i)}s_{\pi(i)}$ is removed from the received  signal. 
 The instantaneous rate of decoding the unicast stream $s_{\pi(k)}$ at user-${\pi(k)}$  is  
\begin{equation}
\label{eq:  DPC unicast rate}
	\begin{aligned}
	&\resizebox{.12 \textwidth}{!} {$R_{\pi(k)}^{\textrm{DPC}}( \mathbf{H},\widehat{\mathbf{H}})=$} \\
	&\resizebox{.032 \textwidth}{!} {$\log_2$}\resizebox{.413 \textwidth}{!} {$ \left(1+\frac{|{\mathbf{h}}_{\pi(k)}^H\mathbf{p}_{\pi(k)}|^2}{\sum_{i<k}|\widetilde{\mathbf{h}}_{\pi(k)}^H\mathbf{p}_{\pi(i)}|^2+\sum_{j> k}|\mathbf{h}_{\pi(k)}^H\mathbf{p}_{\pi(j)}|^2+1}\right).$} 
	\end{aligned}
\end{equation}

Since the BS does not know  the exact channel ${\mathbf{H}}$, precoder design based on instantaneous rate may be overestimated and unachievable at each user. Therefore, a more robust approach is to design precoders according to the Ergodic Rate (ER), which characterizes the long-term rate performance of each stream over all possible joint fading states $\{ \mathbf{H},\widehat{\mathbf{H}}\}$. The  ERs of decoding $s_0$ and $s_{\pi(k)}$ at user-${\pi(k)}$ for conventional DPC-assisted NOUM are defined as\footnote{The achievability of $\overline{R}_{0,\pi(k)}^{\textrm{DPC}}$ and $\overline{R}_{\pi(k)}^{\textrm{DPC}}$ follows the discussion in  Remark 1 of \cite{mao2019beyondDPC}. } 
$
\overline{R}_{0,\pi(k)}^{\textrm{DPC}}\triangleq\mathbb{E}_{\{\mathbf{H},\widehat{\mathbf{H}}\}}\{R_{0,\pi(k)}^{\textrm{DPC}}( \mathbf{H},\widehat{\mathbf{H}})\},
\overline{R}_{\pi(k)}^{\textrm{DPC}}\triangleq\mathbb{E}_{\{ \mathbf{H},\widehat{\mathbf{H}}\}}\{R_{\pi(k)}^{\textrm{DPC}}( \mathbf{H},\widehat{\mathbf{H}})\}
$, respectively.
To ensure $s_0$ is successfully decoded at all users, the ER of the multicast stream $s_0$ should not exceed
\begin{equation}
\label{eq: DPC common ER}
\overline{R}_{0}^{\textrm{DPC}}\triangleq\min\left\{\overline{R}_{0,\pi(k)}^{\textrm{DPC}}\mid k\in \mathcal{K}\right\}. 
\end{equation}

\subsection{ Proposed Dirty Paper Coded Rate-Splitting-assisted NOUM}
\label{sec: DPCRS}
\begin{figure}
		\centering
	\begin{subfigure}[b]{0.45\textwidth}
		\centering
		\includegraphics[width=\textwidth]{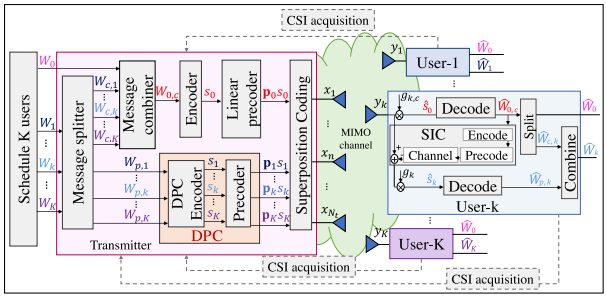}%
		\caption{$K$-user  1-layer Dirty Paper Coded RS. }	
	\end{subfigure}%
	\\
	~
	\centering
	\begin{subfigure}[b]{0.47\textwidth}
		\centering
		\includegraphics[width=\textwidth]{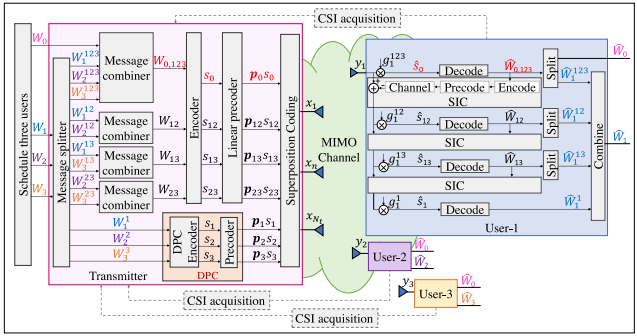}%
		\caption{$3$-user multi-layer Dirty Paper Coded RS.  }
	\end{subfigure}%
	\caption{System architecture of dirty paper coded rate-splitting for non-orthogonal unicast and multicast transmission.}
	\label{fig: NOUM}
		\vspace{-5mm}
\end{figure}

In this work, we aim at exploring larger rate regions of multi-antenna NOUM with partial CSIT by marrying the benefits of DPC and RS. 
The proposed strategies, as illustrated in Fig. \ref{fig: NOUM},  are respectively specified in the following.

\subsubsection{1-DPCRS}
\par  The first strategy we proposed is 1-layer DPCRS (1-DPCRS) in Fig. \ref{fig: NOUM}(a).  The unicast message  $W_k$ intended for user-$k, \forall k\in\mathcal{K}$ is first split into one common part $W_{c,k}$ and one private part $W_{p,k}$. The common parts $W_{c,1},\ldots,W_{c,K}$ of all users are combined with the  multicast stream $W_0$ into the super-common  message $W_c$ and encoded into the super-common  stream $s_c$ to be decoded by all  users. With a certain encoding order $\pi$, the private parts
$W_{p,1},\ldots,W_{p,K}$  are  encoded and precoded by DPC. Denote the stream vector and precoding matrix as  $\mathbf{s}\triangleq[s_c,s_{\pi(1)},\ldots,s_{\pi(K)}]^T$ and $\mathbf{P}\triangleq[\mathbf{p}_c,\mathbf{p}_{\pi(1)},\ldots,\mathbf{p}_{\pi(K)}]$,  the resulting  transmit signal is 
\begin{equation}
\label{eq: transmit signal DPCRS}
\mathbf{x}=\mathbf{P}\mathbf{s}={{\mathbf{p}_c{s}_c}}+{{\sum_{k\in\mathcal{K}}\mathbf{p}_{\pi(k)}{s}_{\pi(k)}}}.
\end{equation}

\par Similarly to the DPC-assisted NOUM, each user-$\pi(k)$ first decodes the super-common stream $s_c$ into $\widehat{W}_c$ by treating  the  interference from all  private streams as noise. The instantaneous rate $R_{c,\pi(k)}^{\textrm{1-DPCRS}}( \mathbf{H},\widehat{\mathbf{H}})$ at user-$\pi(k)$ to decode the super-common  stream $s_c$ is  defined in the same way as the right-hand side of  (\ref{eq: DPC multicast rate}) by replacing $\mathbf{p}_0$ with $\mathbf{p}_c$.
Based on SIC, the decoded super-common message $\widehat{W}_c$ then goes through the process of re-encoding, precoding, and subtracting from the received signal. 
User-$\pi(k)$ then decodes the intended private stream $s_{\pi(k)}$. The instantaneous rate at user-$\pi(k)$ to decode the private stream stream $s_{\pi(k)}$ is  defined in the same way as the right-hand side of (\ref{eq:  DPC unicast rate}), i.e., $R_{\pi(k)}^{\textrm{1-DPCRS}}( \mathbf{H},\widehat{\mathbf{H}})=R_{\pi(k)}^{\textrm{DPC}}( \mathbf{H},\widehat{\mathbf{H}})$. Once $\widehat{W}_c$ and $\widehat{W}_{p,\pi(k)}$ are decoded, user-$\pi(k)$ reconstructs the original multicast and unicast messages by extracting $\widehat{W}_{0}$ and $\widehat{W}_{c,\pi(k)}$ from $\widehat{W}_c$,  and then combines $\widehat{W}_{c,\pi(k)}$ with $\widehat{W}_{p,\pi(k)}$  into  $\widehat{W}_{\pi(k)}$. 
The ERs $\overline{R}_{c,\pi(k)}^{\textrm{1-DPCRS}}$, $\overline{R}_{\pi(k)}^{\textrm{1-DPCRS}}$  of the super-common and private streams are defined as 
$
\overline{R}_{c,\pi(k)}^{\textrm{1-DPCRS}}\triangleq\mathbb{E}_{\{\mathbf{H},\widehat{\mathbf{H}}\}}\{R_{c,\pi(k)}^{\textrm{1-DPCRS}}( \mathbf{H},\widehat{\mathbf{H}})\}$ 
and
$\overline{R}_{\pi(k)}^{\textrm{1-DPCRS}}\triangleq\mathbb{E}_{\{ \mathbf{H},\widehat{\mathbf{H}}\}}\{R_{\pi(k)}^{\textrm{1-DPCRS}}( \mathbf{H},\widehat{\mathbf{H}})\},
$
respectively.
To ensure $s_c$ is successfully decoded by all users, we also have 
$
\overline{R}_{c}^{\textrm{1-DPCRS}}\triangleq\min\left\{\overline{R}_{c,\pi(k)}^{\textrm{1-DPCRS}}\mid k\in \mathcal{K}\right\}. 
$
$\overline{R}_{c}^{\textrm{1-DPCRS}}$ includes the ER of transmitting the multicast stream as well as the ERs of transmitting the common parts of unicast streams.
Denote the ER allocated to the multicast message $W_0$ as $\overline{C}_0$ and the ER allocated to $W_{c, k}$  as $\overline{C}_{k}$, we obtain that
$
\overline{C}_0+\sum_{k\in\mathcal{K}}\overline{C}_{k}=\overline{R}_{c}^{\textrm{1-DPCRS}}.
$
 The   ER of decoding the unicast stream $W_{\pi(k)}$ at user-$\pi(k)$ using 1-DPCRS is 
\begin{equation} 
 \overline{R}_{ \pi(k),tot}^{\textrm{1-DPCRS}}=\overline{C}_{\pi(k)}+\overline{R}_{\pi(k)}^{\textrm{1-DPCRS}}.
 \end{equation}
 
 \subsubsection{M-DPCRS}
 M-DPCRS is an extension of 1-DPCRS  by embracing the generalized RS framework proposed in \cite{mao2019beyondDPC}. The idea is to split the unicast message of each user into more different parts and encode into multiple layers of common streams, each is intended to one  subset of users. The multicast stream is still encoded with some common parts of unicast messages  into the super-common stream to be decoded by all users.  Due to  page limitation,  the system model of M-DPCRS is not specified here. It can be easily traced out from M-DPCRS  in Section II.C of \cite{mao2019beyondDPC} for MISO BC and 1-DPCRS in Fig.  \ref{fig: NOUM}(a) for  multi-antenna NOUM.   Fig. \ref{fig: NOUM}(b) illustrates one example of the proposed  M-DPCRS  for NOUM when $K=3$.

\section{Problem Formulation and Optimization Framework}
\label{sec: problem formulation}
In this section, we formulate the weighted average sum rate  
maximization problem and the specify the corresponding optimization 
framework to solve the problem.
\subsection{Weighted Average Sum Rate  Maximization Problem }
\par  We study the precoder optimization problem at the transmitter with the aim of maximizing the Weighted Ergodic Sum Rate (WESR) of unicast messages subject to the Quality of Service (QoS) rate constraints of multicast and unicast messages. The WESR is defined as $\sum_{k\in\mathcal{K}}u_{k}  \overline{R}_{k,tot}^{\textrm{1-DPCRS}}$, where $u_k$ is the weight for user-$k$. 
We further define the  Average Rate (AR) of decoding the stream $s_i, i\in\{c,k\}$  at user-$k,k\in\mathcal{K}$ for a given channel estimate $\widehat{\mathbf{H}}$ and precoder $\mathbf{P}(\widehat{\mathbf{H}})$ as
\begin{equation}
\label{eq: average rate}
\widehat{R}_{i,k}^{\textrm{1-DPCRS}}(\widehat{\mathbf{H}})\triangleq\mathbb{E}_{\{\mathbf{H}\mid \widehat{\mathbf{H}}\}}\left\{R_{i,k}^{\textrm{1-DPCRS}}(\mathbf{H}, \widehat{\mathbf{H}})\mid \widehat{\mathbf{H}}\right\},
\end{equation}
where $R_{i,k}^{\textrm{1-DPCRS}}(\mathbf{H},\widehat{\mathbf{H}})=R_{k}^{\textrm{1-DPCRS}}(\mathbf{H},\widehat{\mathbf{H}})$  and $\widehat{R}_{i,k}^{\textrm{1-DPCRS}}(\widehat{\mathbf{H}})$ is simplified to $\widehat{R}_{k}^{\textrm{1-DPCRS}}(\widehat{\mathbf{H}})$ when $i=k$. Following \cite{mao2019beyondDPC},  the WESR maximization problem is decomposed into Weighted Average Sum Rate (WASR) maximization problems  to be solved for all possible channel estimates and DPC encoding orders. For a given weight vector $\mathbf{u}=[u_1,\ldots,u_K]$  and a fixed DPC encoding order $\pi$,  the WASR  problem for 1-DPCRS-assisted NOUM  is
\begin{subequations}
	\label{eq: DPCRS b}
	\begin{align}
	&	\max_{ \widehat{\mathbf{c}},\mathbf{{P}}}\,\, \sum_{k\in\mathcal{K}}u_{\pi(k)}(\widehat{C}_{\pi(k)}+\widehat{R}_{\pi(k)}^{\textrm{1-DPCRS}}(\widehat{\mathbf{H}}))\\
	\mbox{s.t.}\,\,
	&\,\, \widehat{C}_{0}+\sum_{k'\in\mathcal{K}}\widehat{C}_{k'}\leq
	 \widehat{R}_{c,k}^{\textrm{1-DPCRS}}(\widehat{\mathbf{H}}), \forall k\in\mathcal{K} \label{c1_DPCRS b}\\
	&  \,\,\widehat{C}_{\pi(k)}+\widehat{R}_{\pi(k)}^{\textrm{1-DPCRS}}(\widehat{\mathbf{H}})\geq R_{\pi(k)}^{th}, \forall k\in\mathcal{K} \label{c2_DPCRS b}\\
		&\,\, \widehat{C}_{0}\geq {R}_{0}^{th}\label{c0_DPCRS NOUM b}\\
	&\,\,	\text{tr}(\mathbf{P}\mathbf{P}^{H})\leq P_{t} \label{c3_DPCRS b}\\	
	&\,\,	\mathbf{\widehat{c}} \geq \mathbf{0}. \label{c4_DPCRS b}
	\end{align}
\end{subequations}
The rate vector $\widehat{\mathbf{c}}=[\widehat{C}_0, \widehat{C}_1,\ldots,\widehat{C}_K]$  for 1-DPCRS-assisted NOUM contains the rates allocated to the multicast message $W_0$ as well as the common parts of the unicast messages $W_{p,1},\ldots,W_{p,K}$ for each $\widehat{\mathbf{H}}$.  It is required to be jointly optimized with the precoders so as to maximize the WASR.   $R_{{\pi(k)}}^{th}$ is the QoS rate constraint of the unicast message $W_{\pi(k)}$ and $R_0^{th}$ is the QoS rate constraint of $W_0$. 

Compared with problem (19) in \cite{mao2019beyondDPC} for 1-DPCRS-assisted MISO BC, the main difference of  problem (\ref{eq: DPCRS b}) comes from constraints (\ref{c1_DPCRS b}) and (\ref{c0_DPCRS NOUM b}) due to the additional multicast message $W_0$ to be transmitted for all users. 
The WASR problem  of  DPC-assisted NOUM is formulated by turning off $\widehat{C}_1,\ldots,\widehat{C}_K$ in (\ref{eq: DPCRS b}). The problem of M-DPCRS-assisted NOUM can be  formulated if readers understand problem (20) in \cite{mao2019beyondDPC} for M-DPCRS-assisted MISO BC and  problem (\ref{eq: DPCRS b})  for 1-DPCRS-assisted NOUM. 

\subsection{Optimization Framework}
\label{sec: algorithm}
\par The formulated  problem  (\ref{eq: DPCRS b}) is stochastic and non-convex. To solve the problem, we extend the optimization framework proposed in \cite{RS2016hamdi}. Specifically, we first transform the original stochastic problem into a deterministic form by using the Sample Average Approximation (SAA) approach. The  approximated deterministic problem is further transformed into an equivalent  Weighted Minimum Mean Square Error (WMMSE) problem, which is then solved by using Alternating Optimization (AO) algorithm. Each step of the optimization framework is further explained in the following.

\par  SAA is first adopted to approximate the stochastic AR (\ref{eq: average rate}) into the corresponding deterministic expression. Assuming that the conditional density $f_{{\mathbf{H}}\mid \widehat{\mathbf{H}}}({\mathbf{H}}\mid \widehat{\mathbf{H}})$ is known at the BS. For a given  $\widehat{\mathbf{H}}$,  the BS  generates a sample of $M$ user channels, indexed by $\mathcal{M}=\{1,\ldots,M\}$ as
\begin{equation}
\label{eq: SAA}
\mathbb{H}^{(M)}\triangleq \{\mathbf{H}^{(m)}=\widehat{\mathbf{H}}+\widetilde{\mathbf{H}}^{(m)}\mid \widehat{\mathbf{H}}, m\in\mathcal{M}\}.
\end{equation}
With the introduced channel sample in (\ref{eq: SAA}) and the strong Law of Large Number (LLN), the ARs $	\widehat{R}_{i,k}^{\textrm{1-DPCRS}}(\widehat{\mathbf{H}})$ specified in equation (\ref{eq: average rate}) is equivalent to 
\begin{equation}
\label{eq: SAF}
\begin{aligned}
\widehat{R}_{i,k}^{\textrm{1-DPCRS}}(\widehat{\mathbf{H}})&= \lim\limits_{M\rightarrow \infty} 
{\frac{1}{M}\sum_{m=1}^{M}R_{i,k}^{\textrm{1-DPCRS}}\left(\mathbf{H}^{(m)},\widehat{\mathbf{H}}\right)}.
\end{aligned}
\end{equation}
Denote 
$\widehat{R}_{i,k}^{{\textrm{1-DPCRS}}^{(M)}}(\widehat{\mathbf{H}})\triangleq{\frac{1}{M}\sum_{m=1}^{M}R_{i,k}^{\textrm{1-DPCRS}}\left(\mathbf{H}^{(m)},\widehat{\mathbf{H}}\right)}$ as the sampled AR with sample size $M$, problem (\ref{eq: DPCRS b}) is transformed into its deterministic form, which is given by
\begin{subequations}
	\label{eq: DPCRS SAA}
	\small
	\begin{align}
	&	\max_{ \widehat{\mathbf{c}},\mathbf{{P}}}\,\, \sum_{k\in\mathcal{K}}u_{\pi(k)}(\widehat{C}_{\pi(k)}+\widehat{R}_{\pi(k)}^{\textrm{1-DPCRS}^{(M)}}(\widehat{\mathbf{H}})) \label{obj: WASR}\\
	\mbox{s.t.}\,\,
	&\,\, \widehat{C}_{0}+\sum_{k'\in\mathcal{K}}\widehat{C}_{k'}\leq
	\widehat{R}_{c,k}^{\textrm{1-DPCRS}^{(M)}}(\widehat{\mathbf{H}}), \forall k\in\mathcal{K} \label{c1_DPCRS SAA}\\
	&  \,\,\widehat{C}_{\pi(k)}+\widehat{R}_{\pi(k)}^{\textrm{1-DPCRS}^{(M)}}(\widehat{\mathbf{H}})\geq R_{\pi(k)}^{th}, \forall k\in\mathcal{K} \label{c2_DPCRS SAA}\\
	&\,\,\textrm{(\ref{c0_DPCRS NOUM b})}, \textrm{(\ref{c3_DPCRS b})}, \textrm{(\ref{c4_DPCRS b})},\nonumber
	\end{align}
\end{subequations}
where the precoder $\mathbf{P}$ and the common stream allocation vector $\widehat{\mathbf{c}}$ are designed over all the $M$ channel samples.

The  approximated deterministic problem (\ref{eq: DPCRS SAA}) is  still non-convex due to the non-convex approximated rate expressions of the common stream and the private streams. To solve the non-convex problem (\ref{eq: DPCRS SAA}), we further extend the WMMSE algorithm proposed in \cite{wmmse08, RS2016hamdi}.  User-$\pi(k)$ employs equalizer $g_{\pi(k)}^i$ to decode  data stream $s_i$.  
The Mean Square Error (MSE) of   stream $s_i, i\in \{c,\pi(k)\}$ at user-$\pi(k)$ is
\begin{equation}
\label{eq:MSE}
\begin{aligned}
\resizebox{.46 \textwidth}{!} {$\varepsilon_{\pi(k)}^i\triangleq\mathbb{E}\{|\widehat{s}_{i}-s_{i}|^{2}\}=|g_{\pi(k)}^i|^2T_{\pi(k)}^i-2\Re\{g_{\pi(k)}^i\mathbf{h}_{\pi(k)}^H\mathbf{p}_{i}\}+1,$}
\end{aligned}
\end{equation}
where $T_{\pi(k)}^{i}\triangleq \sum_{j\in\mathcal{K}\cup\{c\}}|\mathbf{h}_{\pi(k)}^H\mathbf{p}_{j}|^2+1$, if $i=c$ and 
$T_{\pi(k)}^{i}\triangleq |\mathbf{h}_{\pi(k)}^H\mathbf{p}_{\pi(k)}|^2+\sum_{j<k}|\widetilde{\mathbf{h}}_{\pi(k)}^H\mathbf{p}_{\pi(j)}|^2+1,  \textrm{if } i=\pi(k)$.

\par Define the Weighted MSE (WMSE) of decoding $s_i$ at user-$\pi(k)$ as
$
\xi_{\pi(k)}^i( \mathbf{H},\widehat{\mathbf{H}})\triangleq w_{\pi(k)}^i\varepsilon_{\pi(k)}^i-\log_{2}(w_{\pi(k)}^i), 
$
where $w_{\pi(k)}^i$ is the introduced weight for MSE of user-$\pi(k)$. By taking the equalizers and weights as optimization variables, the Rate--WMMSE relationship for an instantaneous channel realization is established as
\resizebox{.45\textwidth}{!} {$
\xi_{i,\pi(k)}^{\textrm{MMSE}}( \mathbf{H},\widehat{\mathbf{H}})\triangleq\min_{w_{\pi(k)}^i,g_{\pi(k)}^i}\xi_{\pi(k)}^i( \mathbf{H},\widehat{\mathbf{H}})=1-R_{i,\pi(k)}^{\textrm{1-DPCRS}}( \mathbf{H},\widehat{\mathbf{H}}).
$}
By defining $w_{\pi(k)}^{i,(m)},g_{\pi(k)}^{i,(m)}$ as the weights and equalizers associated with the $m$th channel realization in $\mathbb{H}^{(M)}$, the relationship for an instantaneous channel realization  is then extended to the average Rate-WMMSE relationships over $\mathbb{H}^{(M)}$ as
\begin{equation}
\label{eq: rate-wmmse average}
\resizebox{.4\textwidth}{!} {$\begin{aligned}
\widehat{\xi}_{i,\pi(k)}^{^{(M)}}(\widehat{\mathbf{H}})&\triangleq{{\frac{1}{M}\sum_{m=1}^{M}\left(\min_{w_{\pi(k)}^{i,(m)},g_{\pi(k)}^{i,(m)}}\xi_{\pi(k)}^i( \mathbf{H}^{(m)},\widehat{\mathbf{H}})\right)}}\\
&=1-{\widehat{R}_{i,\pi(k)}^{\textrm{1-DPCRS}^{(M)}}}(\widehat{\mathbf{H}}).
\end{aligned}$}
\end{equation}
${\widehat{\xi}_{i,\pi(k)}^{(M)}}( \widehat{\mathbf{H}})$ is simplified to $\widehat{\xi}_{\pi(k)}^{{(M)}}( \widehat{\mathbf{H}})$ when $i=\pi(k)$. 
Based on (\ref{eq: rate-wmmse average}), problem (\ref{eq: DPCRS SAA}) is equivalently transformed   into the WMMSE problem, which is given by
\begin{subequations}
	\label{eq: RS WMMSE}
	\small
	\begin{align}
	&	\min_{ \mathbf{{P}},\widehat{\mathbf{x}},\mathbf{w},\mathbf{g}}\,\, \sum_{k\in\mathcal{K}}u_{\pi(k)}(\widehat{X}_{\pi(k)}+\widehat{\xi}_{\pi(k)}^{{(M)}}( \widehat{\mathbf{H}})) \label{obj: WMMSE}\\
	\mbox{s.t.}\,\,
	&\,\, \widehat{X}_{0}+\sum_{k'\in\mathcal{K}}\widehat{X}_{k'}+1\geq \widehat{\xi}_{c,k}^{{(M)}}( \widehat{\mathbf{H}}), \forall k\in\mathcal{K} \label{c1_RS WMMSE}\\
	&  \,\,\widehat{X}_{\pi(k)}+\widehat{\xi}_{\pi(k)}^{{(M)}}( \widehat{\mathbf{H}})\leq 1-R_{\pi(k)}^{th}, \forall k\in\mathcal{K} \label{c2_RS WMMSE}\\
	&  \,\,\widehat{X}_{0}\leq -R_{0}^{th}, \forall k\in\mathcal{K} \label{c5_RS WMMSE}\\
	&\,\,		\text{tr}(\mathbf{P}\mathbf{P}^{H})\leq P_{t}  \label{c4_RS WMMSE}\\
	&\,\,\widehat{\mathbf{x}} \leq \mathbf{0},  \label{c3_RS WMMSE} 
	\end{align}
\end{subequations}
where $\widehat{\mathbf{x}}=[\widehat{X}_0, \widehat{X}_1,\ldots,\widehat{X}_K]$ is the transformation of $\widehat{\mathbf{c}}$ satisfying  $\widehat{\mathbf{x}}=-\widehat{\mathbf{c}}$.   \resizebox{.3\textwidth}{!}{$\mathbf{w}=\{w_{\pi(k)}^{i,(m)} |i\in\{c,\pi(k)\}, k\in\mathcal{K},m\in\mathcal{M}\}$}  and \resizebox{.3\textwidth}{!}{$\mathbf{g}=\{g_{\pi(k)}^{i,(m)}|i\in\{c,\pi(k)\},   k\in\mathcal{K},m\in\mathcal{M}\}$} are the MSE weights and equalizers, respectively.

\setlength{\textfloatsep}{7pt}	
\begin{algorithm}[t!]
	\textbf{Initialize}: $n\leftarrow0$, $\mathbf{P}$, $\mathrm{WASR}^{[n]}$\;
	\Repeat{$|\mathrm{WASR}^{[n]}-\mathrm{WASR}^{[n-1]}|\leq \epsilon$}{
		$n\leftarrow n+1$\;
		$\mathbf{P}^{[n-1]}\leftarrow \mathbf{P}$\;
		update $\mathbf{g}$ and $\mathbf{w}$  by $\mathbf{g}^{\star}(\mathbf{P}^{[n-1]})$ and $\mathbf{w}^{\star}(\mathbf{P}^{[n-1]})$ specified in (\ref{eq: equalizer update}) and (\ref{eq: weight update}), respectively\; 
		update $(\mathbf{P},\widehat{\mathbf{x}})$ by solving (\ref{eq: RS WMMSE final}) using the updated $\mathbf{w}, \mathbf{g}$;	
	}	
	\caption{WMMSE-based AO algorithm}
	\label{WMMSE algorithm}				
\end{algorithm}

Problem (\ref{eq: RS WMMSE}) is block-wise convex with respect to each block of $\mathbf{w}$, $\mathbf{g}$ and $(\mathbf{{P}},\widehat{\mathbf{x}})$ by fixing other two blocks, which motivates us to use AO algorithm to solve the problem.  At each iteration $[n]$, for  given $\mathbf{w}^{[n-1]}$ and $(\mathbf{{P}}^{[n-1]},\widehat{\mathbf{x}}^{[n-1]})$, the  solution $\mathbf{g}^{[n]}=\mathbf{g}^{\star}(\mathbf{{P}}^{[n-1]})$ of   (\ref{eq: RS WMMSE}) is
\begin{equation}
\label{eq: equalizer update}
\resizebox{.48\textwidth}{!}{$\mathbf{g}^{\star}(\mathbf{{P}}^{[n-1]})\triangleq\left\{\mathbf{p}_i^H\mathbf{h}_{\pi(k)}^{(m)}(T_{\pi(k)}^{i,(m)})^{-1}\mid i\in\{c,\pi(k)\}, k\in\mathcal{K},m\in\mathcal{M}\right\}$}
\end{equation}    
and the  solution $\mathbf{w}^{[n]}=\mathbf{w}^{\star}(\mathbf{{P}}^{[n-1]})$ of   (\ref{eq: RS WMMSE}) for  given $\mathbf{g}^{[n-1]}$ and $(\mathbf{{P}}^{[n-1]},\widehat{\mathbf{x}}^{[n-1]})$ is 
\begin{equation}
\label{eq: weight update}
\resizebox{.48\textwidth}{!}{$\mathbf{w}^{\star}(\mathbf{{P}}^{[n-1]})\triangleq\left\{\frac{T_{\pi(k)}^{i,(m)}}{T_{\pi(k)}^{i,(m)}-|(\mathbf{h}_{\pi(k)}^{(m)})^H\mathbf{p}_i|^2}\mid i\in\{c,\pi(k)\}, k\in\mathcal{K},m\in\mathcal{M}\right\}$},
\end{equation}  
where $T_{\pi(k)}^{i,(m)}$ in (\ref{eq: equalizer update}) and (\ref{eq: weight update}) is calculated based on the $m$th channel sample in  $\mathbb{H}^{(M)}$. The solutions of weights and equalizers in  (\ref{eq: equalizer update}) and (\ref{eq: weight update}) satisfy the   Karush-Kuhn-Tucker (KKT) conditions of  (\ref{eq: RS WMMSE}). Substituting (\ref{eq: equalizer update}) and (\ref{eq: weight update}) back to   (\ref{eq: RS WMMSE}), the optimization problem becomes:
\begin{subequations}
	\label{eq: RS WMMSE final}
	\begin{align}
	&	\min_{ \mathbf{{P}},\widehat{\mathbf{x}}}\,\, \sum_{k\in\mathcal{K}}u_{\pi(k)}(\widehat{X}_{\pi(k)}+\widehat{\xi}_{\pi(k)}^{\textrm{1-DPCRS}}( \widehat{\mathbf{H}})) \label{obj: WMMSE final}\\
	\mbox{s.t.}\,\,
	&\,\, \widehat{X}_{0}+\sum_{k'\in\mathcal{K}}\widehat{X}_{k'}+1\geq \widehat{\xi}_{c,k}^{\textrm{1-DPCRS}}( \widehat{\mathbf{H}}), \forall k\in\mathcal{K} \label{c1_RS WMMSE final}\\
	&  \,\,\widehat{X}_{\pi(k)}+\widehat{\xi}_{\pi(k)}^{\textrm{1-DPCRS}}( \widehat{\mathbf{H}})\leq 1-R_{\pi(k)}^{th}, \forall k\in\mathcal{K} \label{c2_RS WMMSE final}\\
	&  \,\,\textrm{(\ref{c5_RS WMMSE})}, \textrm{(\ref{c4_RS WMMSE})},  \textrm{(\ref{c3_RS WMMSE})},\nonumber
	\end{align}
\end{subequations}
	where  
$	\widehat{\xi}_{i,\pi(k)}^{\textrm{1-DPCRS}}( \widehat{\mathbf{H}})\triangleq\Omega_{\pi(k)}^{i}+\bar{t}_{\pi(k)}^i-2\Re\left\{(\bar{\mathbf{f}}_{\pi(k)}^i)^H\mathbf{p}_{i}\right\}+\bar{w}_{\pi(k)}^i-\bar{\nu}_{{\pi(k)}}^i
$
and it is simplified to $\widehat{\xi}_{\pi(k)}^{\textrm{1-DPCRS}}( \widehat{\mathbf{H}})$ when $i=\pi(k)$. $\Omega_{\pi(k)}^{i}=	\sum_{j\in\mathcal{K}\cup\{c\}}\mathbf{p}_{j}^H\bar{\Psi}_{\pi(k)}^i\mathbf{p}_{ j}$ if   $i=c$ and $\Omega_{\pi(k)}^{i}=\mathbf{p}_{j}^H\bar{\Psi}_{\pi(k)}^i\mathbf{p}_{ j}+\sum_{j< k}\mathbf{p}_{\pi(j)}^H\bar{\Phi}_{\pi(k)}^i\mathbf{p}_{\pi(j)}$ if   $i=\pi(k)$.
$\bar{\Psi}_{\pi(k)}^i, $ $ \bar{\Phi}_{\pi(k)}^i, \bar{t}_{\pi(k)}^i,  \bar{\mathbf{f}}_{\pi(k)}^i, \bar{w}_{\pi(k)}^i,  \bar{\nu}_{{\pi(k)}}^i$ are constants (or constant vectors/matrices) averaged over a sample of $M$ user channels, i.e., $\bar{w}_{\pi(k)}=\frac{1}{M}\sum_{m=1}^M{w}_{\pi(k)}^{i,(m)}$. Their corresponding values in each channel instance $(m)$ are updated as
\begin{equation}
\begin{aligned}
& {t}_{\pi(k)}^{i,(m)}=w_{\pi(k)}^{i,(m)}\left|g_{\pi(k)}^{i,(m)}\right|^2, \quad
 {\Psi}_{\pi(k)}^{i,(m)}={t}_{\pi(k)}^{i,(m)}\mathbf{h}_{\pi(k)}^{(m)}(\mathbf{h}_{\pi(k)}^{(m)})^H,\\ 
& {\nu}_{{\pi(k)}}^{i,(m)}=\log_2\left(w_{\pi(k)}^{i,(m)}\right),\quad {\Phi}_{\pi(k)}^{i,(m)}={t}_{\pi(k)}^{i,(m)}\widetilde{\mathbf{h}}_{\pi(k)}^{(m)}(\widetilde{\mathbf{h}}_{\pi(k)}^{(m)})^H,\\
&\mathbf{f}_{\pi(k)}^{i,(m)}=w_{\pi(k)}^{i,(m)}\mathbf{h}_{\pi(k)}^{(m)}(g_{\pi(k)}^{i,(m)})^H.
\end{aligned}
\end{equation}

\par Problem (\ref{eq: RS WMMSE final})  is a standard Quadratically Constrained Quadratic Program (QCQP), which can be solved using interior-point methods \cite{boyd2004convex}. Therefore, we obtain the  AO algorithm  specified in Algorithm \ref{WMMSE algorithm}. The weights $\mathbf{w}$, equalizers $\mathbf{g}$, precoders and common rate vectors $(\mathbf{P},\widehat{\mathbf{x}})$ are updated iteratively until  the WASR of the system  $\mathrm{WASR}^{[n]}$  converges. The convergence proof of Algorithm \ref{WMMSE algorithm} follows   \cite{RS2016hamdi, mao2019beyondDPC}, which is not specified here. By using the same method, we could also obtain the formulated problem and the corresponding optimization framework for DPC and M-DPCRS-assisted NOUM.

\section{Numerical Results}
\label{sec: numerical results}
\par In this section, we evaluate the performance of the proposed 1-DPCRS and M-DPCRS strategies for  NOUM. CVX toolbox \cite{grant2008cvx} is adopted to tackle problem (\ref{eq: RS WMMSE final}) that requires to be solved by the  interior-point method. The exact channel $\mathbf{h}_k$ and the channel estimation error $\widetilde{\mathbf{h}}_k$ have i.i.d. complex Gaussian entries  drawn from the distributions $\mathcal{CN}(0,\sigma_k^2)$, $\mathcal{CN}(0,\sigma_{e,k}^2)$, respectively and $\sigma_{e,k}^2=\sigma_k^2P_t^{-\alpha}$.  The sample size of SAA method is $M=1000$. The WESR is obtained by averaging  WASR over 100 channel realizations.  The precoder initialization  for Algorithm \ref{WMMSE algorithm} follows the methods in \cite{mao2019beyondDPC}. The QoS rate constraint of the multicast stream is $R_0^{th}=0.5$ bit/s/Hz. SNR is 20 dB. 
 We compare the following eight transmission strategies in the results.   ``1-DPCRS"  and ``M-DPCRS"   are the strategies we proposed in Section \ref{sec: DPCRS}. ``DPC" is the strategy described in Section \ref{sec: DPC}. ``Generalized RS", ``1-layer RS", ``SC--SIC", ``SC--SIC per group" and ``MU--LP"  are the linearly precoded strategies proposed in \cite{mao2019TCOM} for multi-antenna NOUM.

 \begin{figure}[t!]
 	 		\centering
 	\begin{subfigure}[b]{0.22\textwidth}
 		\centering
 		\includegraphics[width=\textwidth]{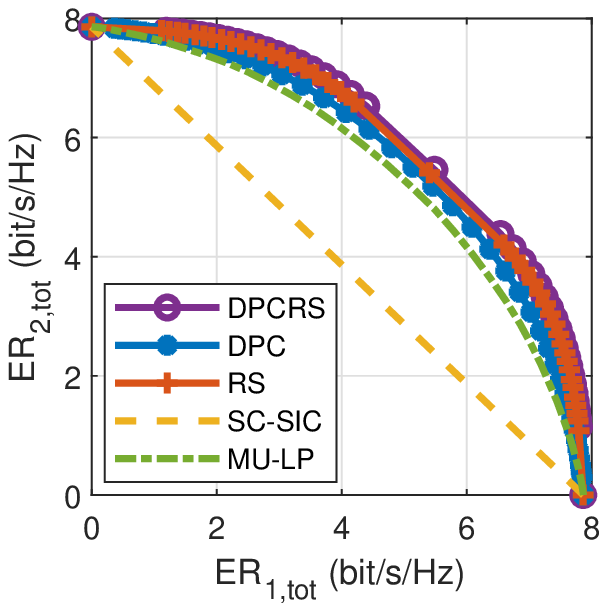}%
 		\caption{$N_t=4, \sigma_2^2=1$}
 	\end{subfigure}%
 	~
 	\begin{subfigure}[b]{0.22\textwidth}
 		\centering
 		\includegraphics[width=\textwidth]{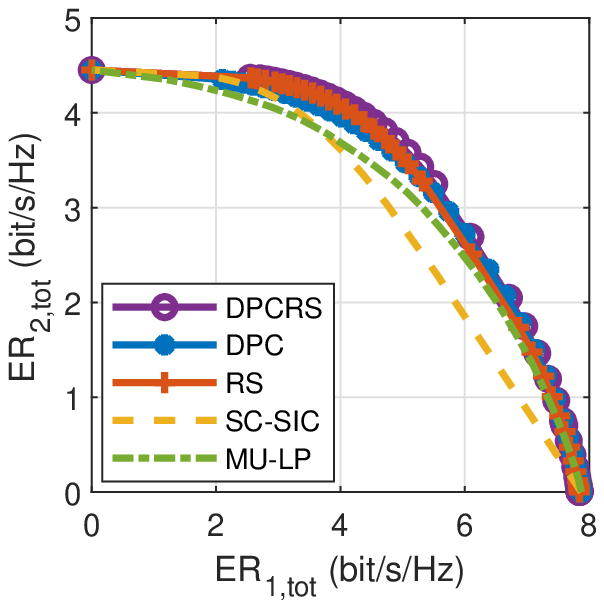}%
 		\caption{$N_t=4, \sigma_2^2=0.09$}
 	\end{subfigure}%
 	~\\
 	\begin{subfigure}[b]{0.22\textwidth}
 		\centering
 		\includegraphics[width=\textwidth]{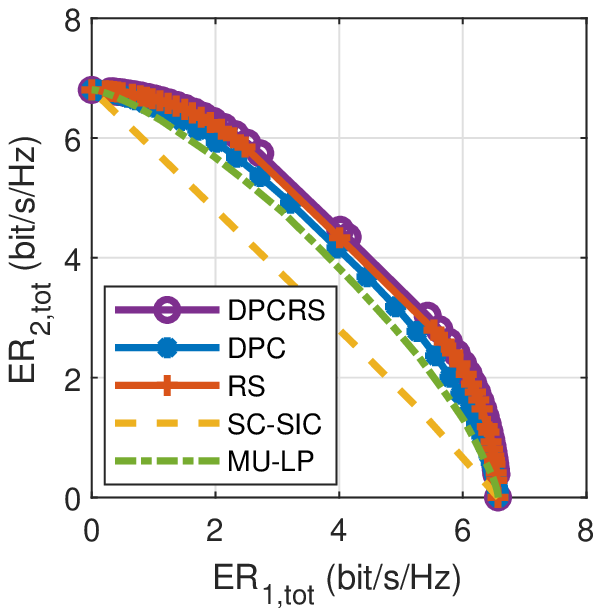}%
 		\caption{$N_t=2, \sigma_2^2=1$}
 	\end{subfigure}%
 	~
 	\begin{subfigure}[b]{0.22\textwidth}
 		\centering
 		\includegraphics[width=\textwidth]{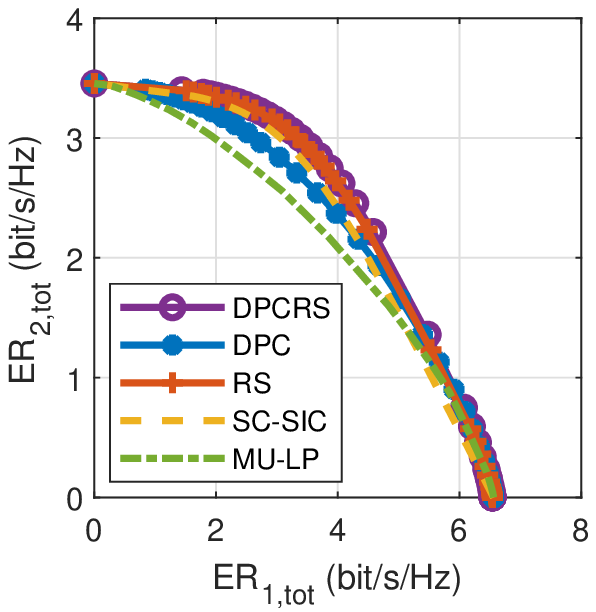}%
 		\caption{$N_t=2, \sigma_2^2=0.09$}
 	\end{subfigure}%
 	\caption{Ergodic rate region comparison of different strategies with partial CSIT for multi-antenna NOUM, averaged over 100 random channel realizations, SNR=20 dB, K = 2, $\alpha=0.6$, $\sigma_1^2=1$, $R_0^{th}$= 0.5 bit/s/Hz.}
 	\label{fig: Ergodic snr20 alpha06 NOUM}
 \end{figure}

Fig. \ref{fig: Ergodic snr20 alpha06 NOUM} illustrates the two-user ER region comparison of all strategies. When $K=2$, we use the term “DPCRS” to represent both M-DPCRS
and 1-DPCRS and use the term “RS” to represent both generalized RS and 1-layer RS since M-DPCRS and  generalized RS  respectively reduces to 1-DPCRS and 1-layer RS.  In all subfigures, DPCRS maintains the largest rate region compared with the rate regions of all other strategies. Interesting, we found that the existing linearly precoded RS studied in \cite{mao2019TCOM}, benefiting from its robustness towards partial CSIT, outperforms DPC in most of cases. 

\begin{figure}[t!]
	\centering
	\begin{subfigure}[b]{0.22\textwidth}
		\centering
		\includegraphics[width=\textwidth]{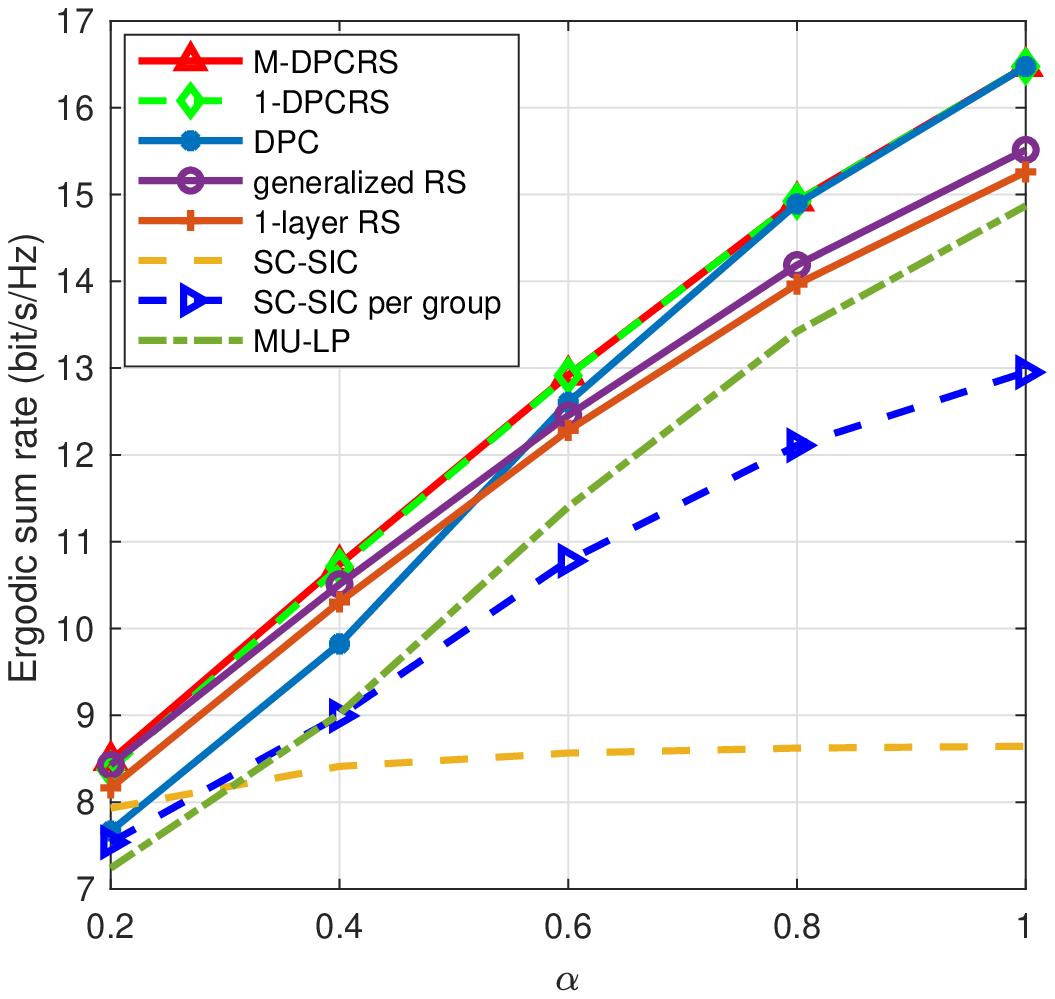}
		\caption{$\mathbf{r}_k^{th}=[0.1,\ldots,0.5]$ bit/s/Hz, $N_t=4$, $\sigma_1^2=\sigma_2^2=\sigma_3^2=1$.}
	\end{subfigure}%
	~ 
	\begin{subfigure}[b]{0.22\textwidth}
		\centering
		\includegraphics[width=\textwidth]{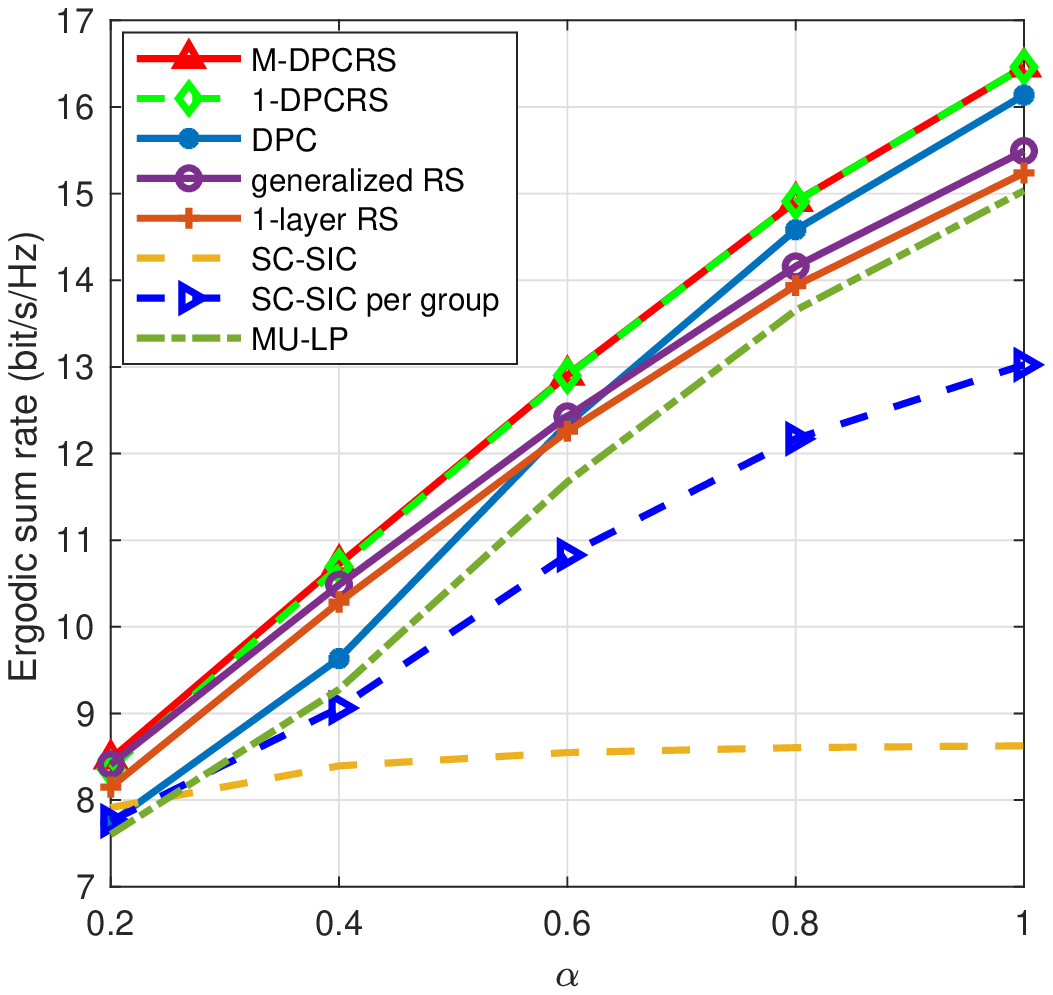}
		\caption{$R_k^{th}=0$,  $N_t=4$, $\sigma_1^2=\sigma_2^2=\sigma_3^2=1$.}
	\end{subfigure}%
	~ \\
	\begin{subfigure}[b]{0.22\textwidth}
		\centering
		\includegraphics[width=\textwidth]{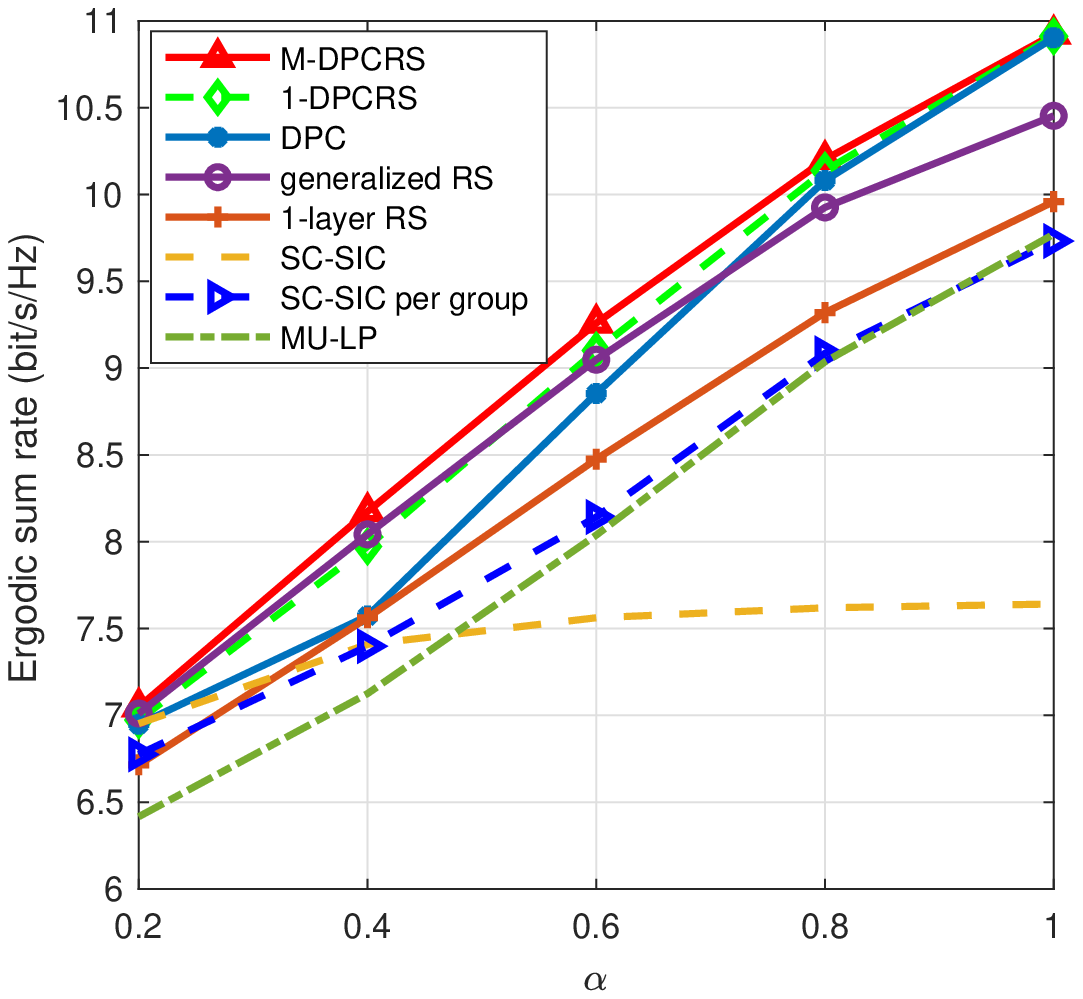}
		\caption{$R_k^{th}=0$,  $N_t=2$, $\sigma_1^2=\sigma_2^2=\sigma_3^2=1$.}
	\end{subfigure}%
	~ 
	\begin{subfigure}[b]{0.22\textwidth}
		\centering
		\includegraphics[width=\textwidth]{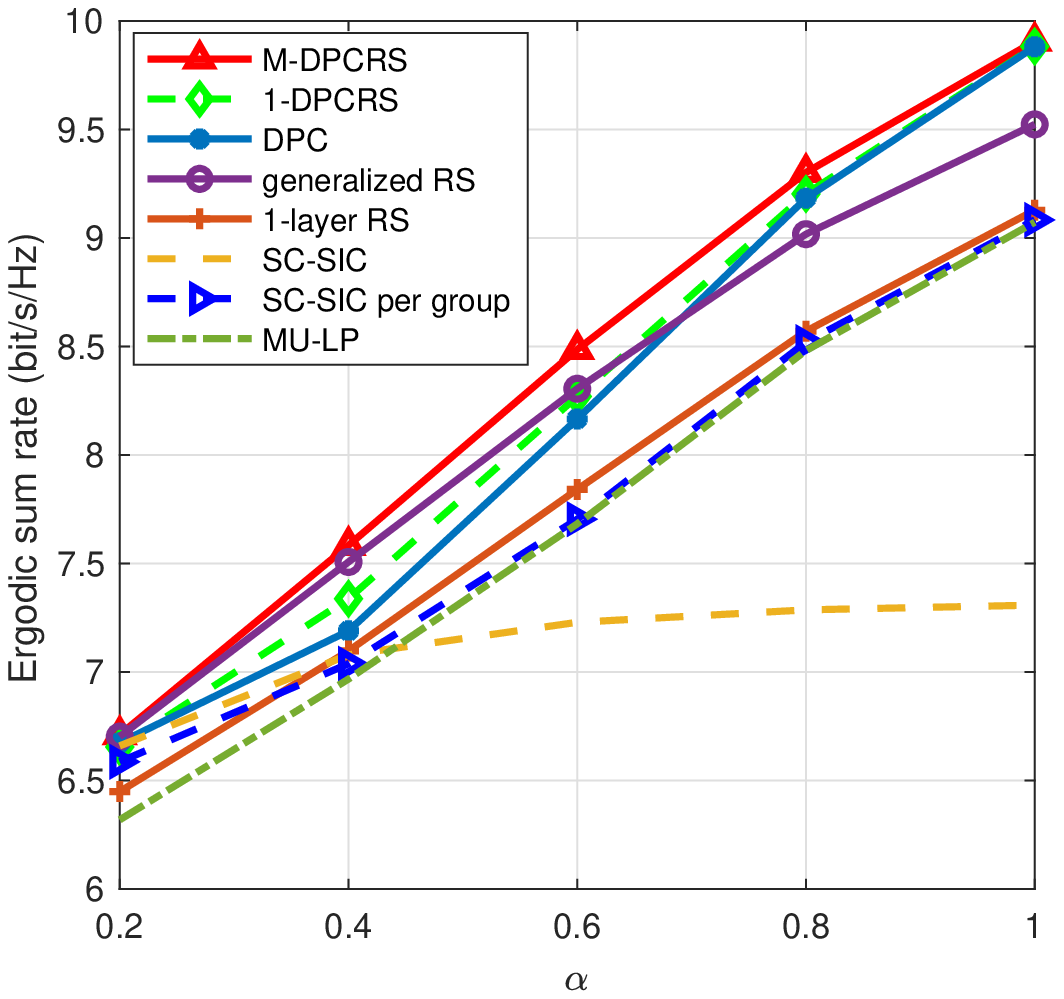}%
		\caption{$R_k^{th}=0$,  $N_t=2$, $\sigma_1^2=\sigma_2^2=1,\sigma_3^2=0.3$.}
	\end{subfigure}
	\caption{Ergodic sum rate versus CSIT inaccuracy $\alpha$  comparison of different strategies for multi-antenna NOUM, averaged over 100 random channel realizations, $K=3$,  SNR = 20 dB, $R_0^{th}=0.5$ bit/s/Hz.}
	\label{fig: SRwithAlpha bias11 NOUM}
\end{figure}

In the three-user case, we study the Ergodic Sum Rate (ESR, i.e., $u_k=1, \forall k\in\mathcal{K}$) comparison versus CSIT accuracy in Fig. \ref{fig: SRwithAlpha bias11 NOUM}.  Overall, M-DPCRS achieves the highest  ESR  with explicit ESR improvement over DPC, MU--LP and SC--SIC-assisted strategies. Linearly precoded RS strategies (generalized RS and 1-layer RS) outperform non-linear DPC especially in the region with strong CSIT inaccuracy.

\section{Conclusion}
\label{sec: conclusion}
\par In this work, we propose a novel strategy, namely, Dirty Paper Coded Rate-Splitting (DPCRS) 
that incorporates RS with DPC to assess the  rate region of multi-antenna non-orthogonal unicast and multicast transmission with partial CSIT. By splitting the unicast messages of each user into common and private parts,  using DPC to encode the private parts, jointly encoding the multicast message with the common parts of the unicast messages, DPCRS is able to partially decode the interference and partially treat inference as noise, further restrain the interference between multicast and unicast messages as well as  the multi-user interference  among unicast messages. Numerical results show that linearly precoded RS-assisted NOUM is able to achieve larger rate region than DPC-assisted NOUM  but with a much lower hardware and computational complexities. The proposed DPCRS-assisted NOUM outperforms all existing strategies. It is more robust to CSIT inaccuracies, network loads and user deployments.

\bibliographystyle{IEEEtran}
\bibliography{reference}
\end{document}